\documentclass[aps,prd,a4paper,twocolumn,longbibliography,amsmath,showpacs,superscriptaddress,nofootinbib,preprintnumbers]{revtex4-1}
\pdfoutput=1
\usepackage{amssymb,amsmath,latexsym,mathrsfs}
\usepackage[sort&compress]{natbib}
\usepackage{graphicx,subfigure}
\usepackage{epsfig}
\usepackage{varioref,xr-hyper}
\usepackage{color}
\usepackage{multirow}
\usepackage{array}
\usepackage{hyperref}
\usepackage{wasysym}
\usepackage{color}
\usepackage{float}
\usepackage{xcolor}
\usepackage[utf8]{inputenc}
\usepackage[T1]{fontenc}
\definecolor{ao}{rgb}{0.0, 0.5, 0.0}
\hypersetup{colorlinks,linkcolor={blue},citecolor={blue},urlcolor={ao}}

\begin{document}





\title{Probing the cold nature of dark matter}







\author{Weiqiang Yang}
\email{d11102004@163.com}
\affiliation{Department of Physics, Liaoning Normal University, Dalian, 116029, People's Republic of China}

\author{Supriya Pan}
\email{supriya.maths@presiuniv.ac.in}
\affiliation{Department of Mathematics, Presidency University, 86/1 College Street, Kolkata 700073, India}
\affiliation{Institute of Systems Science, Durban University of Technology, PO Box 1334, Durban 4000, Republic of South Africa}

\author{Eleonora Di Valentino}
\email{e.divalentino@sheffield.ac.uk}
\affiliation{School of Mathematical and Physical Sciences, University of Sheffield, Hounsfield Road, Sheffield S3 7RH, United Kingdom}

\author{Olga Mena}
\email{omena@ific.uv.es}
\affiliation{Instituto de F\'{i}sica Corpuscular (CSIC-Universitat de Val\`{e}ncia), E-46980 Paterna, Spain} 

\author{David F. Mota}
\email{d.f.mota@astro.uio.no}
\affiliation{Institute of Theoretical Astrophysics, University of Oslo, P.O. Box 1029 Blindern, N-0315 Oslo, Norway}

\author{Subenoy Chakraborty}
\email{schakraborty.math@gmail.com}
\affiliation{Department of Mathematics, Brainware University, Barasat, West Bengal 700125, India}
\affiliation{Shinawatra University, 99 Moo 10, Bangtoey, Samkhok, Pathum Thani 12160 Thailand}
\affiliation{INTI International University, Persiaran Perdana BBN, Putra Nilai, 71800 Nilai, Malaysia}


\begin{abstract}
\noindent A pressureless dark matter component fits well with several cosmological observations. However, there are indications that cold dark matter may encounter challenges in explaining observations at small scales, particularly at galactic scales. Observational data suggest that dark matter models incorporating a pressure component could provide solutions to these small-scale problems.  In this work, we investigate the possibility that present-day dark matter may result from a decaying non-cold dark matter sector transitioning into the dark energy sector. As the sensitivity of astronomical surveys rapidly increases, we explore an interacting scenario between dark energy and non-cold dark matter, where dark energy has a constant equation of state ($w_{\rm de}$), and dark matter, being non-cold, also has a constant (non-zero) equation of state ($w_{\rm dm}$).  
Considering the phantom and quintessence nature of dark energy, characterized by its equation of state, we separately analyze interacting phantom and interacting quintessence scenarios. We constrain these scenarios using Cosmic Microwave Background (CMB) measurements and their combination with external probes, such as DESI-BAO and PantheonPlus. From our analyses, we find that a very mild preference for non-cold dark matter cannot be excluded based on the employed datasets. Additionally, for some datasets, there is a pronounced preference for the presence of an interaction at more than 95\% confidence level (CL). Moreover, when the dark energy equation of state lies in the phantom regime, the $S_8$ tension can be alleviated. 
This study suggests that cosmological models incorporating a non-cold dark matter component should be considered as viable scenarios with novel phenomenological implications, as reflected in the present work.

\end{abstract}

\maketitle
\section{Introduction}

The $\Lambda$-Cold Dark Matter ($\Lambda$CDM) model has achieved great success over the years due to its simplicity and its agreement with a variety of cosmological probes. Within this simple cosmological framework, our universe is primarily dominated by a positive cosmological constant and a cold dark matter (CDM) component, where the CDM sector comprises nearly 28\% of the total energy budget of the universe, while the cosmological constant $\Lambda$ accounts for approximately 68\% of the total energy density~\cite{Planck:2018vyg}. 
Despite the success of the $\Lambda$CDM model, the nature of its two main components, namely the cosmological constant $\Lambda$ and CDM, remains unknown. Moreover, recent advancements in observational cosmology suggest that estimates of some key cosmological parameters within the minimal $\Lambda$CDM framework, derived from Cosmic Microwave Background (CMB) observations by Planck~\cite{Planck:2018vyg}, differ significantly—at multiple standard deviations—from those obtained through other astronomical observations. One notable example is the estimation of the Hubble constant by Planck (within the $\Lambda$CDM paradigm)~\cite{Planck:2018vyg}, which is in $5\sigma$ tension with the measurement obtained by the SH0ES (Supernova $H_0$ for the Equation of State of Dark Energy) collaboration~\cite{Riess:2021jrx}. 
Furthermore, the estimated value of the $S_8$ parameter, defined as the combination of the amplitude of the matter power spectrum $\sigma_8$ with the present-day matter density $\Omega_{m0}$,  
$S_8 = \sigma_8 \sqrt{\Omega_{m0}/0.3}$,  
as measured by Planck within the $\Lambda$CDM framework, differs from the values extracted from other astronomical missions, such as weak gravitational lensing and galaxy clustering~\cite{DES:2021bvc,DES:2021vln,Asgari:2020wuj,DAmico:2020kxu,Asgari:2019fkq,Joudaki:2019pmv,DAmico:2019fhj,Ivanov:2019pdj}. 
These discrepancies indicate that a revision of the $\Lambda$CDM cosmology is essential to address the mismatches in cosmological parameter estimates across different astronomical surveys. As a result, a variety of cosmological models beyond the standard $\Lambda$CDM paradigm have been proposed~\cite{Abdalla:2022yfr,DiValentino:2021izs,Perivolaropoulos:2021jda,Schoneberg:2021qvd}. Nevertheless, the search for a cosmological model that can satisfactorily resolve all the existing tensions within the standard $\Lambda$CDM framework is still ongoing. 

In most cosmological models, the equation of state of the dark matter (DM) component is typically set to zero, implying that dark matter is assumed to be pressureless. This assumption is justified by the abundance of the CDM component in the universe and by the fact that setting the DM equation of state to zero reduces degeneracies in the parameter space. However, there is ongoing debate regarding the possibility of a nonzero equation of state for DM, which would suggest the existence of non-cold DM in the universe. This possibility has been explored in several works in the literature~\cite{Muller:2004yb,Kumar:2012gr,Armendariz-Picon:2013jej,Kopp:2018zxp,Ilic:2020onu,Naidoo:2022rda}. 
Since the fundamental nature of both DM and dark energy (DE) remains unclear, and given that recent observations suggest the need for a revision of $\Lambda$CDM cosmology, it is natural to consider a more complete picture in which the equation of state of DM is allowed to be nonzero and determined by the observational data. From a theoretical standpoint, one may speculate that the current abundance of CDM could result from a decaying non-cold DM component transitioning into DE over the history of the universe. This hypothesis implies an interaction mechanism between DE and non-cold DM that could explain the present abundance of CDM. 
Interacting dark sector scenarios have been widely investigated over the past several years due to their potential to address various cosmological puzzles, ranging from the cosmic coincidence problem~\cite{Amendola:1999er,Chimento:2003iea,Cai:2004dk,Pavon:2005yx,Huey:2004qv,delCampo:2008sr,delCampo:2008jx} to the current cosmological tensions~\cite{Kumar:2017dnp,DiValentino:2017iww,Yang:2018euj,Pan:2019gop,Pourtsidou:2016ico,An:2017crg,Kumar:2019wfs,Shah:2024rme}. Numerous authors have explored these scenarios in detail (see, e.g.,~\cite{Amendola:1999er,Billyard:2000bh,Amendola:2002bs,Mangano:2002gg,Cai:2004dk,Pavon:2005yx,Huey:2004qv,Amendola:2006dg,Barrow:2006hia,He:2008tn,Quartin:2008px,Koyama:2009gd,Majerotto:2009np,Chimento:2009hj,Gavela:2009cy,Gavela:2010tm,Clemson:2011an,Yang:2014gza,Yang:2014vza,Yang:2014hea,Salvatelli:2014zta,Li:2015vla,Nunes:2016dlj,Feng:2016djj,vandeBruck:2016jgg,Costa:2016tpb,Kumar:2016zpg,vandeBruck:2016hpz,Erdem:2016hqw,Pan:2016ngu,Kumar:2017dnp,DiValentino:2017iww,Mifsud:2017fsy,VanDeBruck:2017mua,Sharov:2017iue,vandeBruck:2017idm,vonMarttens:2018iav,Martinelli:2019dau,Pan:2019jqh,Yang:2019uog,DiValentino:2019ffd,DiValentino:2019jae,Lucca:2020zjb,Pan:2020bur,Pan:2020zza,DiValentino:2020leo,Paliathanasis:2021egx,Gao:2021xnk,Yang:2021oxc,Johnson:2020gzn,Kumar:2021eev,Wang:2021kxc,Guo:2021rrz,Kang:2021osc,Bonilla:2021dql,Gariazzo:2021qtg,Jesus:2020tby,Potting:2021bje,Nunes:2022bhn,Tsedrik:2022cri,Chatzidakis:2022mpf,Zhai:2023yny,vanderWesthuizen:2023hcl,Pan:2023mie,Forconi:2023hsj,Sa:2023coi,Benisty:2024lmj,Halder:2024uao,Silva:2024ift,Giare:2024ytc,Giare:2024smz,Aboubrahim:2024spa,Pooya:2024wsq,Halder:2024gag,Li:2024qso,Sabogal:2024yha,Ashmita:2024ueh,Benetti:2024dob,Halder:2024aan,Lima:2024wmy,Li:2025owk,Sabogal:2025mkp,Tsedrik:2025cwc,Zhai:2025hfi,Silva:2025hxw,Pan:2025qwy}; see also reviews~\cite{Bolotin:2013jpa,Wang:2016lxa,Wang:2024vmw}). 
Following this perspective, in the present work, we explore a non-gravitational interaction scenario between DE and non-cold DM, aiming to determine whether a decaying non-cold DM model is favored by current observational data.

The article is organized as follows. In~\autoref{sec-2}, we describe the basic equations governing the interaction between non-cold dark matter and dark energy at both the background and perturbation levels. Then, in~\autoref{sec-datasets}, we present the observational datasets and the statistical methodology employed in our analysis. In~\autoref{sec-results}, we discuss the results obtained from our study. Finally, in~\autoref{sec-summary}, we summarize our main findings and provide concluding remarks.

\section{Interaction between non-cold dark matter and dark energy}
\label{sec-2}

We consider the spatially flat Friedmann-Lema\^{i}tre-Robertson-Walker (FLRW) line element, given by  

\begin{equation}
ds^2 = -dt^2 + a^2 (t) (dx^2 + dy^2 + dz^2),
\end{equation}  
where $a(t)$ is the expansion scale factor of the universe, and $(t, x, y, z)$ are the comoving coordinates. We assume that the gravitational sector of the universe is well described by General Relativity, and within the matter sector, there exists a non-gravitational interaction between two dark fluids, namely dark energy and non-cold dark matter.  
The conservation equations for dark energy and non-cold dark matter in the presence of their interaction are given by  

\begin{eqnarray}
&&\dot{\rho}_{\rm de} + 3(1+w_{\rm de}) H \rho_{\rm de} = Q (t), \label{cons-de}\\
&&\dot{\rho}_{\rm dm} + 3(1+w_{\rm dm}) H \rho_{\rm dm} = - Q (t), \label{cons-wdm}
\end{eqnarray}  
where an overhead dot denotes the derivative with respect to cosmic time, and $H$ is the Hubble rate of the FLRW universe, which satisfies the constraint  

\begin{equation}
3H^2 = 8 \pi G (\rho_{b} + \rho_{r} + \rho_{\nu} + \rho_{\rm dm} + \rho_{\rm de}),
\end{equation}  
where $\rho_b$, $\rho_r$, $\rho_{\nu}$, $\rho_{\rm dm}$, and $\rho_{\rm de}$ denote the energy densities of baryons, radiation, neutrinos (assuming one massive neutrino with a fixed mass of $0.06$ eV and two massless neutrinos), non-cold dark matter with a constant equation-of-state (EoS) parameter, and dark energy with a constant EoS parameter $w_{\rm de}$, respectively.  
A crucial term in the above equations is the interaction function $Q (t)$, which governs the transfer of energy and/or momentum between the dark sectors. The sign of $Q(t)$ determines the direction of energy transfer: for $Q (t) > 0$, energy flows from DM to DE, while for $Q (t) < 0$, energy flows from DE to DM.  Note that eqns. (\ref{cons-de}) and (\ref{cons-wdm}) can alternatively be written as, $\dot{\rho}_{\rm de} + 3 \left(1+ w_{\rm de}^{\rm eff} \right) H \rho_{\rm de} =0$, and $\dot{\rho}_{\rm dm} + 3 \left(1+ w_{\rm dm}^{\rm eff} \right) H \rho_{\rm dm} =0$,  
where $w_{\rm de}^{\rm eff} \equiv w_{\rm de} - Q(t)/(3H \rho_{\rm de})$, $w_{\rm dm}^{\rm eff} \equiv w_{\rm dm} + Q(t)/(3H \rho_{\rm dm})$ respectively denote the effective EoS for DE and DM. One can clearly notice that, in the presence of an interaction between these dark sectors, the effective EoS parameters of the dark components could be dynamical even if the individual nature of the EoS parameter remains non-dynamical. In fact, one can view this interacting scenario as an effective prescription of two non-interacting dark fluids enjoying dynamical EoS. 
Once the interaction function is specified, the dynamics of the universe can be determined either analytically or numerically. In the present work, we consider a well-known interaction function \cite{He:2008tn,Gavela:2009cy,Yang:2014gza,Yang:2014vza,Yang:2014hea,Wang:2016lxa,Costa:2016tpb,DiValentino:2019ffd,DiValentino:2019jae,Pan:2020zza,Lucca:2020zjb,Pan:2022qrr,Hou:2022rvk,Giare:2024smz,Benisty:2024lmj,Li:2024qso,Feng:2025mlo} 

\begin{equation}
Q = 3 H \xi \rho_{\rm de},
\end{equation}  
where $\xi$ is the coupling parameter of the interaction function, assumed to be constant. In an expanding universe (i.e., $H > 0$), a positive coupling parameter ($\xi > 0$) implies energy transfer from DM to DE, while a negative coupling parameter ($\xi < 0$) signifies energy transfer from DE to DM.\footnote{It should be noted here that one can equally consider a different interaction model which is either proportional to the energy density of DM, i.e. $Q = 3 H \xi \rho_{\rm dm}$~\cite{Amendola:2006dg,Wang:2016lxa}, or which could involve the energy densities of both DM and DE, e.g. $Q = 3 H \xi (\rho_{\rm dm} + \rho_{\rm de})$~\cite{Wang:2016lxa} (one can consider a more general form such as $Q = 3 H \xi_1 \rho_{\rm dm} + \xi_2 \rho_{\rm de}$~\cite{Quartin:2008px,Pan:2012ki,Pan:2016ngu,Sharov:2017iue}) or $Q = 3 H \xi \rho_{\rm dm} \rho_{\rm de} \times (\rho_{\rm dm} + \rho_{\rm de})^{-1}$~\cite{Li:2013bya}.  
All of them are phenomenologically very appealing.  
A comparison between the outcomes of these interaction models will shed further light on this particular interacting set-up where DM enjoys a non-cold EoS. }  
For this interaction function, solving the conservation equations~\eqref{cons-de} and~\eqref{cons-wdm} leads to explicit solutions for the energy densities of the dark components:  

\begin{equation}
\rho_{\rm de} = \rho_{\rm de,0}\;a^{-3(1+w_{\rm de}-\xi)},\label{eqn-rho-DM} 
\end{equation}  

\begin{eqnarray}
\rho_{\rm dm} &=& \rho_{\rm dm,0}\;a^{-3(1+w_{\rm dm})}  +  \rho_{\rm de,0}\; \left(\frac{\xi}{w_{\rm dm}+\xi-w_{\rm de}} \right) \nonumber \\ 
&& \times  \Bigg[a^{-3(1+w_{\rm dm})}-a^{-3(1+w_{\rm de}-\xi)} \Bigg], \label{eqn-rho-DE}
\end{eqnarray}  
where $\rho_{\rm dm,0}$ and $\rho_{\rm de,0}$ denote the present values of $\rho_{\rm dm}$ and $\rho_{\rm de}$, respectively.  
Notably, in the absence of interaction (i.e., $\xi = 0$), the standard evolution equations for DM and DE are recovered, corresponding to a non-interacting cosmological scenario. Thus, using equations~\eqref{eqn-rho-DM} and~\eqref{eqn-rho-DE}, along with the Hubble equation, one can, in principle, determine the background evolution (either analytically or numerically) of the universe.

Once the evolution at the background level is determined, it is essential to investigate the dynamics of the interacting scenario at the perturbation level. To explore this, we consider the perturbed metric of the FLRW spacetime in a general gauge, given by~\cite{Mukhanov:1990me,Ma:1995ey,Malik:2008im}:

\begin{eqnarray}
ds^2 = a^2(\tau) \Bigg[ -(1+2\phi)d\tau^2 + 2\partial_i B d\tau dx^i \nonumber\\ 
+ \bigg\{(1-2\psi)\delta_{ij} + 2\partial_i\partial_j E \bigg\} dx^i dx^j \Bigg],
\label{eq:perturbed-metric}
\end{eqnarray}
where $\phi$, $B$, $\psi$, and $E$ are gauge-dependent scalar quantities representing metric perturbations, and $\tau$ denotes the conformal time.
The four-velocity of a given fluid component $A$ (where $A = \{\rm de, \rm dm\}$ represents either DE or DM) is given by~\cite{Valiviita:2008iv,Yang:2014gza}:

\begin{eqnarray}
u^{\mu}_A = a^{-1} (1 - \phi, \partial^i v_A),
\end{eqnarray}
where $v_A$ is the peculiar velocity potential of fluid $A$. In Fourier space $k$, it is related to the volume expansion $\theta_A$ via:

\begin{eqnarray}
\theta_A = -k^2 (v_A + B).
\end{eqnarray}
A general energy-momentum transfer in the context of an interaction between two fluids can be decomposed as~\cite{Valiviita:2008iv,Yang:2014gza}:

\begin{equation}
Q^{\mu}_A = \widetilde{Q}_A u^{\mu} + F^{\mu}_A,
\end{equation}
where
\begin{equation}
\widetilde{Q}_A = Q_A + \delta Q_A, \quad F^{\mu}_A = a^{-1} (0, \partial^i f_A).
\end{equation}
Here, $Q_A$ represents the background interaction rate, while $f_A$ is the momentum transfer potential. The perturbed energy-momentum transfer four-vector is then decomposed as~\cite{Valiviita:2008iv,Yang:2014gza}:

\begin{eqnarray}
Q^A_0 &=& -a [Q_A (1 + \phi) + \delta Q_A], \\
Q^A_i &=& a \partial_i [Q_A (v + B) + f_A].
\end{eqnarray}
Now, the evolution equations for the dimensionless density perturbation $\delta_A = \delta\rho_A / \rho_A$ and the velocity perturbation ($\theta_A$) can be written as (avoiding the anisotropic stress of the fluids)~\cite{Valiviita:2008iv,Yang:2014gza,Yang:2014vza,Yang:2014hea}:
\begin{align}
\delta'_A+3\mathcal{H}(c^2_{sA}-w_A)\delta_A
+9\mathcal{H}^2(1+w_A)(c^2_{sA}-c^2_{aA})\frac{\theta_A}{k^2}
& \nonumber \\  + (1+w_A)\theta_A
-3(1+w_A)\psi'+(1+w_A)k^2(B-E')
\nonumber \\
=\frac{a}{\rho_A}(-Q_A\delta_A+\delta Q_A)
+\frac{aQ_A}{\rho_A}\Bigg[\phi +3\mathcal{H}(c^2_{sA}-c^2_{aA})\frac{\theta_A}{k^2}\Bigg],
\label{eq:general-deltaA} \\
\theta'_A+\mathcal{H}(1-3c^2_{sA})\theta_A-\frac{c^2_{sA}}{(1+w_A)}k^2\delta_A
-k^2\phi 
& \nonumber \\
=\frac{a}{(1+w_A)\rho_A}\Bigg[(Q_A\theta-k^2f_A)-(1+c^2_{sA})Q_A\theta_A \Bigg],
\label{eq:general-thetaA}
\end{align}
where the prime corresponds to differentiation with respect to the conformal time $\tau$; $\mathcal{H} = a^{\prime}/a$ is the conformal Hubble rate; $c^2_{aA}$ is the adiabatic sound speed of the fluid $A$, defined as $c^2_{aA}=p'_A/\rho'_A = w_A + w'_A/(\rho'_A/\rho_A)$; $c^2_{sA}$ is the physical sound speed of the fluid $A$ in the rest frame. For a barotropic fluid, they are equal, i.e. $c^2_{sA} = c^2_{aA}$, and for constant $w_A$, which we have considered in this article, $c^2_{sA} = w_A$. Therefore, the sign of $w_A$ is important, because for negative $w_A$, $c^2_{sA}$ becomes imaginary, which is unphysical, and hence this quantity should be non-negative~\cite{Valiviita:2008iv}. 
Now, in the synchronous gauge, i.e. $\phi =B=0$, $\psi =\eta $, and $k^{2}E=-h/2-3\eta$, in which $h$, $\eta$ denote the metric perturbations (see \cite{Ma:1995ey} for more details), one can write down the above equations explicitly for DE and DM components as follows: 
\begin{widetext}
\begin{eqnarray}
&&\delta _{\rm de}^{\prime } =-(1+w_{\rm de})\left( \theta _{\rm de}+\frac{h^{\prime }}{2}%
\right) -3\mathcal{H}(c_{\rm s, de}^{2}-w_{\rm de})\left[ \delta _{\rm de}+3\mathcal{H}%
(1+w_{\rm de})\frac{\theta _{\rm de}}{k^{2}}\right]
+9\mathcal{H}^2\xi(c_{\rm s, de}^{2}-w_{\rm de})\frac{\theta _{\rm de}}{k^{2}}, \\
&&\theta _{\rm de}^{\prime } =-\mathcal{H}(1-3c_{\rm s,de}^{2})\theta _{\rm de}+\frac{%
c_{\rm s, de}^{2}}{(1+w_{\rm de})}k^{2}\delta _{\rm de}+3\mathcal{H}\xi\left[ \frac{%
\theta _{\rm dm}-(1+c_{\rm s, de}^{2})\theta _{\rm de}}{1+w_{\rm de}}\right] , \\
&&\delta _{\rm dm}^{\prime } =-(1+w_{\rm dm})\left( \theta _{\rm dm}+\frac{h^{\prime }}{2}%
\right) -3\mathcal{H}(c_{\rm s, dm}^{2}-w_{\rm dm})\left[ \delta _{\rm dm}+3\mathcal{H}%
(1+w_{\rm dm})\frac{\theta _{\rm dm}}{k^{2}}\right] \notag \\
&&~~~~~~~~~~+3\mathcal{H}\xi\frac{\rho _{\rm de}}{\rho _{\rm dm}}\left[ -\delta _{\rm dm}+\delta_{\rm de}+3\mathcal{H}%
(c_{\rm s, dm}^{2}-w_{\rm dm})\frac{\theta _{\rm dm}}{k^{2}}\right] , \\
&&\theta _{\rm dm}^{\prime } =-\mathcal{H}(1-3c_{\rm s, dm}^{2})\theta _{\rm dm}+\frac{%
c_{\rm s, dm}^{2}}{(1+w_{\rm dm})}k^{2}\delta _{\rm dm}+3\mathcal{H}\xi\frac{\rho _{\rm de}}{\rho _{\rm dm}}\left[ \frac{%
\theta _{\rm dm}-(1+c_{\rm s, dm}^{2})\theta _{\rm dm}}{1+w_{\rm dm}}\right]. 
\end{eqnarray}
\end{widetext}
Following the earlier notations,  $c^2_{\rm s, dm}$, $c^2_{\rm s, de}$ respectively denotes the physical sound speeds of non-cold dark matter and dark energy in their respective rest frames. And as already mentioned, we should be careful about the sign of both $c^2_{\rm s, dm}$, $c^2_{\rm s, de}$. Following \cite{Valiviita:2008iv}, we fix $c^2_{\rm s, de} = 1$, and on the other hand, for the sake of simplicity, and since dark matter is responsible for the formation of large-scale structures in the universe, the sound speed of non-cold dark matter is set to $c^2_{\rm s, dm} = 0$.  We close this section with the choice of the parameter space of the underlying interacting scenario, because the early-time instabilities of the interacting models are sensitive to the parameter space. As discussed in~\cite{Gavela:2009cy}, the stability of an interacting model can be assessed through the analysis of the {\it doom factor}, which in our case becomes $d \equiv -Q/[3H (1+w_{\rm de}) \rho_{\rm de}] = - \xi /(1+w_{\rm de})$. The early-time stability of the interaction model requires $d \leq 0$~\cite{Gavela:2009cy}. 
Hereafter, based on this doom factor analysis ~\cite{Gavela:2009cy}, we test two models (with different parameter priors) against observational datasets:

\begin{itemize}
    \item[(i)] $\xi \leq 0$ and $w_{\rm de} < -1$, labeled as {\bf IWDM-DEp} ($p$ stands for phantom dark energy).
    \item[(ii)] $\xi \geq 0$ and $w_{\rm de} > -1$, labeled as {\bf IWDM-DEq} ($q$ corresponds to quintessence dark energy).
\end{itemize}
Note that IWDM (Interacting Dark Matter with equation-of-state W) follows the earlier nomenclature introduced in~\cite{Pan:2022qrr}.

\section{Observational datasets}
\label{sec-datasets} 

In this section, we describe the observational datasets and the numerical methodology adopted to constrain the interacting scenarios. We utilize the following cosmological probes:

\begin{itemize}

\item {\bf CMB}: We use Cosmic Microwave Background (CMB) observations from the Planck 2018 team~\cite{Planck:2018vyg,Aghanim:2019ame}. Specifically, we consider the CMB temperature and polarization angular power spectra {\it plikTTTEEE+lowl+lowE}.  

\item {\bf DESI-BAO}: We incorporate the BAO measurements from the DESI collaboration~\cite{DESI:2024mwx} in the redshift range $0.1<z<4.16$. These measurements are based on the clustering of multiple tracers, including the Bright Galaxy Sample (BGS), the Luminous Red Galaxy (LRG) sample, the Emission Line Galaxy (ELG) sample, the combined LRG+ELG sample, quasars, and the Lyman-$\alpha$ forest. This dataset serves as our baseline BAO sample.

\item {\bf PantheonPlus}: Type Ia supernovae (SNe Ia) provide crucial distance measurements for probing the background evolution of the universe, particularly the equation of state of dark energy. In this analysis, we utilize SN data from the Pantheon+ sample~\cite{Scolnic:2021amr}, which consists of 1701 light curves of 1550 spectroscopically confirmed SNe Ia from eighteen different surveys. 

\end{itemize}

To constrain the interacting scenarios {\bf IWDM-DEp} and {\bf IWDM-DEq}, we have  
modified the \texttt{CosmoMC} package~\cite{Lewis:2002ah}, a publicly available cosmological analysis framework (available at \url{http://cosmologist.info/cosmomc/}). This package supports the Planck 2018 likelihood~\cite{Aghanim:2019ame} and employs a convergence diagnostic based on the Gelman and Rubin criterion~\cite{Gelman:1992zz}.  
Both interacting scenarios involve nine free parameters. Six of these parameters are identical to those in the standard flat $\Lambda$CDM model, while the additional three free parameters are $w_{\rm dm}$, $\xi$, and $w_{\rm de}$. Table~\ref{tab:priors} summarizes the priors imposed on the free parameters considered in this analysis.

\begin{table*}
\begin{center}
\renewcommand{\arraystretch}{1.4}
\begin{tabular}{|c@{\hspace{1 cm}}|@{\hspace{1 cm}}c|@{\hspace{1 cm}}c|}
\hline
\textbf{Parameter}                    & \textbf{Prior (IWDM-DEp)} & \textbf{Prior (IWDM-DEq)} \\
\hline\hline
$\Omega_{b} h^2$             & $[0.005,0.1]$   &  $[0.005,0.1]$\\
$\Omega_{dm} h^2$             & $[0.01,0.99]$   & $[0.01,0.99]$\\
$\tau$                       & $[0.01,0.8]$  & $[0.01,0.8]$ \\
$n_s$                        & $[0.5, 1.5]$ & $[0.5, 1.5]$ \\
$\log[10^{10}A_{s}]$         & $[2.4,4]$ & $[2.4,4]$ \\
$100\theta_{MC}$             & $[0.5,10]$ & $[0.5,10]$ \\ 
$w_{\rm dm}$                & $[0, 1]$  & $[0, 1]$ \\
$\xi$                        & $[-1,0]$   & $[0, 1]$\\
$w_{\rm de}$                & $[-2,-1]$  & $[-1,0]$ \\

\hline
\end{tabular}
\end{center}
\caption{Priors imposed on the free parameters of the proposed cosmological scenarios for the statistical analysis.}
\label{tab:priors}
\end{table*}

\begingroup                                                                                                                     
\squeezetable                                                                                                                   
\begin{center}                                                                                                                  
\begin{table*} 
\begin{tabular}{cccccccc}      
\hline\hline      
Parameters & CMB &  CMB+DESI & CMB+PantheonPlus & CMB+DESI+PantheonPlus  &  \\ \hline
$\Omega_{\rm dm} h^2$ & $    0.133_{-    0.013-    0.015}^{+    0.006+    0.018}$ & $    0.133_{-    0.012-    0.015}^{+    0.007+    0.016}$  
& $ 0.138_{-    0.005-    0.014}^{+    0.010+    0.012}$  & $    0.137_{-    0.005-    0.015}^{+    0.010+    0.013}$ \\

$\Omega_b h^2$ & $    0.02230_{-    0.00015-    0.00030}^{+    0.00015+    0.00030}$ &  $    0.02230_{-    0.00015-    0.00030}^{+    0.00015+    0.00029}$  & $    0.02226_{-    0.00015-    0.00030}^{+    0.00015+    0.00030}$  & $    0.02233_{-    0.00015-    0.00031}^{+    0.00015+    0.00030}$ \\

$100\theta_{MC}$ & $    1.03998_{-    0.00055-    0.0011}^{+    0.00065+    0.0010}$ & $    1.03998_{-    0.00058-    0.0010}^{+    0.00058+    0.0010}$ & $    1.03967_{-    0.00052-    0.00088}^{+    0.00045+    0.00096}$  & $    1.03985_{-    0.00054-    0.00088}^{+    0.00042+    0.00096}$ \\

$\tau$ & $    0.0546_{-    0.0081-    0.015}^{+    0.0075+    0.016}$ &  $    0.0555_{-    0.0077-    0.015}^{+    0.0078+    0.016}$  & $    0.0548_{-    0.0083-    0.015}^{+    0.0075+    0.017}$  & $    0.0568_{-    0.0085-    0.015}^{+    0.0077+    0.017}$ \\

$n_s$ & $    0.9740_{-    0.0044-    0.0088}^{+    0.0044+    0.0088}$ &   $    0.9749_{-    0.0041-    0.0080}^{+    0.0041+    0.0080}$  & $    0.9729_{-    0.0044-    0.0084}^{+    0.0043+    0.0086}$ & $    0.9769_{-    0.0039-    0.0076}^{+    0.0039+    0.00752}$ \\

${\rm{ln}}(10^{10} A_s)$ & $    3.055_{-    0.016-    0.031}^{+    0.016+    0.032}$ &  $    3.056_{-    0.016-    0.032}^{+    0.016+    0.032}$  & $    3.056_{-    0.016-    0.031}^{+    0.016+    0.034}$   & $    3.057_{-    0.018-    0.032}^{+    0.016+    0.034}$  \\

$w_{\rm dm}$ & $ <0.00035, <0.00081$ &  $ <0.00042, <0.00085 $  & $ <0.00033, <0.00076$  & $  0.00050_{-    0.00048}^{+    0.00015}, < 0.00110 $ \\

$w_{\rm de}$ & $ -1.56_{-    0.35-    0.43}^{+    0.20+    0.49}$ &  $ -1.137_{-    0.062}^{+    0.077}, > -1.261 $  & $   > -1.054, > -1.096 $   & $ >-1.055, > -1.096 $ \\

$\xi$  & $ >-0.048  > -0.086 $ &  $ > -0.062 > -0.098  $  & $   -0.063_{-    0.039-    0.046}^{+    0.019+    0.051}$ & $ -0.063_{-    0.038-    0.045}^{+    0.019+    0.051}$ \\

$\Omega_{m0}$ & $    0.224_{-    0.070-    0.083}^{+    0.029+    0.107}$ & $    0.314_{-    0.027-    0.043}^{+    0.022+    0.046}$  & $    0.361_{-    0.019-    0.042}^{+    0.025+    0.038}$  & $    0.348_{-    0.016-    0.038}^{+    0.024+    0.034}$ \\

$\sigma_8$ & $    0.889_{-    0.085-    0.15}^{+    0.085+    0.15}$ & $    0.778_{- 0.045-    0.074}^{+    0.044+    0.076}$   & $    0.738_{-    0.042-    0.054}^{+    0.023+    0.066}$  & $    0.733_{-    0.042-    0.055}^{+    0.022+    0.067}$ \\

$H_0$ & $ > 78.0 $ & $   70.6_{-    1.8-    3.1}^{+    1.5+    3.3}$ & $   66.76_{-    0.83-    1.6}^{+    0.82+    1.7}$  & $   67.80_{- 0.67-    1.3}^{+    0.68+    1.3}$  \\

$S_8$ & $    0.755_{-    0.038-    0.063}^{+    0.034+    0.069}$ &   $0.794_{-    0.020-    0.036}^{+    0.019+    0.037}$ &  $    0.809_{-    0.021-    0.038}^{+    0.019+    0.041}$  & $    0.788_{-    0.020-    0.034}^{+    0.016+    0.036}$ \\

$r_{\rm{drag}}$ & $  147.21_{-    0.31-    0.61}^{+    0.31+    0.62}$ & $147.29_{-    0.27-    0.54}^{+    0.27+    0.53}$ &  $147.14_{-    0.30-    0.60}^{+    0.30+    0.59}$  & $  147.45_{-    0.25-    0.49}^{+    0.25+    0.50}$ \\

\hline\hline     
\end{tabular}            
\caption{Observational constraints on the interacting scenario {\bf IWDM-DEp} using various datasets.}
\label{tab:IWDM-DEp-1}   
\end{table*}               
\end{center}                 
\endgroup  
\begin{figure*}
\centering
\includegraphics[width=0.85\textwidth]{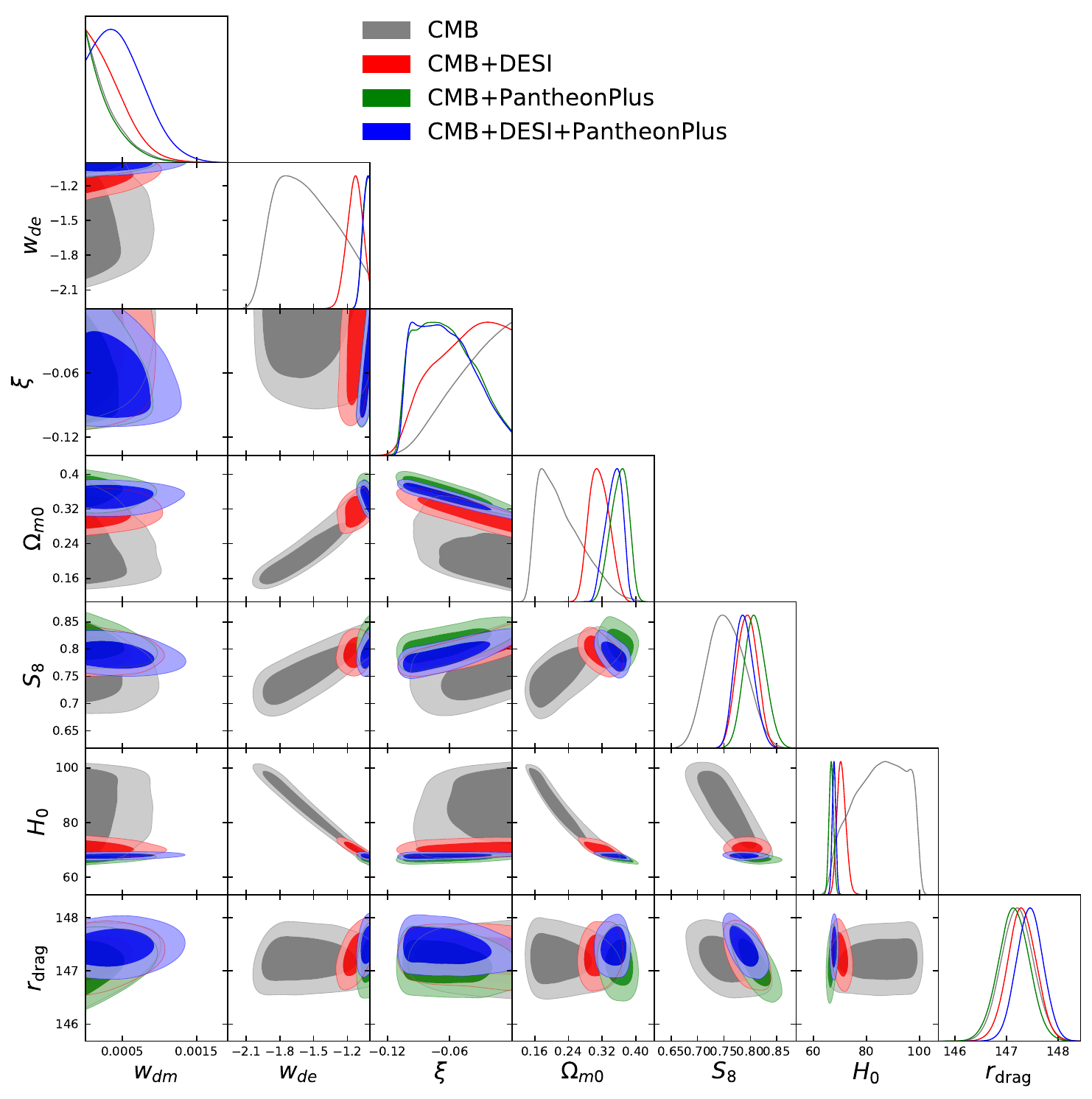}
    \caption{One-dimensional marginalized posterior distributions and two-dimensional joint contours for the interacting scenario {\bf IWDM-DEp}, considering different combinations of cosmological measurements.}
    \label{fig:IDWM-DEp-1}
\end{figure*}

\begingroup                                                                                                                     
\squeezetable     
\begin{center}                                           
\begin{table*}                             
\begin{tabular}{cccccccc}                                       
\hline\hline                                  
Parameters & CMB & CMB+DESI & CMB+PantheonPlus & CMB+DESI+PantheonPlus \\ \hline
$\Omega_{\rm dm} h^2$ & $<0.080\,<0.113$ &   $    0.061_{-0.028-    0.052}^{+    0.034+    0.049}$ & $    0.086_{-    0.012-    0.056}^{+    0.035+    0.039}$  & $    0.090_{-    0.010-    0.045}^{+    0.027+    0.031}$ \\

$\Omega_b h^2$ & $    0.02226_{-    0.00015-    0.00029}^{+    0.00015+    0.00029}$ & $    0.02230_{-    0.00015-    0.00030}^{+    0.00015+    0.00030}$ & $    0.02224_{-    0.00014-    0.00031}^{+    0.00016+    0.00029}$ & $    0.02233_{-    0.00016-    0.00032}^{+    0.00016+    0.00030}$  \\

$100\theta_{MC}$ & $    1.0446_{-    0.0037-    0.0042}^{+    0.0020+    0.0048}$ & $    1.0445_{-    0.0028-    0.0036}^{+    0.0016+    0.0042}$  & $    1.0428_{-    0.0023-    0.0027}^{+    0.0007+    0.0040}$  & $    1.0426_{-    0.0017-    0.0021}^{+    0.0007+    0.0030}$  \\

$\tau$ & $    0.0547_{- 0.0076-    0.015}^{+    0.0076+    0.016}$ & $    0.0561_{-    0.0080-    0.015}^{+    0.0074+    0.016}$  & $    0.0548_{-    0.0076-    0.015}^{+ 0.0075+    0.016}$  & $    0.0571_{-    0.0083-    0.015}^{+    0.0072+    0.017}$ \\

$n_s$ & $    0.9732_{-    0.0043-    0.0083}^{+    0.0042+    0.0086}$ & $    0.9754_{-    0.0039-    0.0078}^{+    0.0040+    0.0078}$   & $    0.9730_{-    0.0043-    0.0084}^{+    0.0043+    0.0086}$ & $    0.9774_{-    0.0038-    0.0079}^{+    0.0039+    0.0078}$  \\

${\rm{ln}}(10^{10} A_s)$ & $    3.056_{-    0.016-    0.031}^{+    0.016+    0.032}$ & $    3.057_{-    0.015-    0.031}^{+    0.016+    0.032}$  & $    3.056_{-    0.016-    0.032}^{+    0.016+    0.032}$   & $    3.058_{-    0.017-    0.032}^{+    0.015+    0.035}$ \\

$w_{\rm dm}$ & $ <0.00031, < 0.00074 $ & $ < 0.00046 < 0.00093 $  & $ < 0.00032 < 0.00076$   
& $ 0.00052_{-    0.00050}^{+    0.00015}, < 0.00116 $  \\

$w_{\rm de}$ & $  >-0.885, >-0.747$ & $ > -0.905, > -0.824 $  & $   -0.880_{-    0.091-    0.12}^{+    0.047+    0.13}$  & $   -0.899_{-    0.072-    0.10}^{+    0.043+    0.11}$  \\

$\xi$ & $  < 0.2653, < 0.2657 $ & $    0.146_{-    0.067-    0.12}^{+    0.078+    0.12}$  & $ < 0.120 < 0.224$  & $ <0.094, < 0.177$ \\

$\Omega_{m0}$ & $    0.18_{-    0.13-    0.14}^{+    0.06+    0.16}$ & $    0.173_{-    0.069-    0.12}^{+    0.066+    0.11}$  & $    0.245_{-    0.030-    0.13}^{+    0.080+    0.09}$ & $    0.248_{-    0.024-    0.099}^{+    0.060+    0.071}$  \\

$\sigma_8$ & $    1.65_{-    0.98-    1.1}^{+    0.21+    2.0}$ & $    1.50_{-    0.70-    0.85}^{+    0.16+    1.42}$  & $    1.10_{-    0.35-    0.43}^{+    0.03+    0.79}$  & $    1.00_{-    0.22-    0.27}^{+    0.04+    0.47}$ \\

$H_0$ & $   69.2_{-    2.9-    7.0}^{+    4.3+    6.5}$ & $   70.2_{-    1.6-    2.8}^{+    1.4+    3.1}$  & $   66.59_{-    0.89-    1.7}^{+    0.83+    1.7}$ & $   67.65_{-    0.70-    1.3}^{+    0.70+    1.4}$ \\

$S_8$ & $    1.08_{-    0.27-    0.30}^{+    0.08+    0.50}$ & $    1.03_{-    0.21-    0.26}^{+    0.06+    0.40}$   & $    0.94_{-    0.12-    0.15}^{+    0.02+    0.27}$  &  $    0.886_{-    0.082-    0.10}^{+    0.022+    0.17}$  \\

$r_{\rm{drag}}$ & $  147.16_{-    0.30-    0.60}^{+    0.30+    0.60}$ & $  147.33_{-    0.27-    0.52}^{+    0.26+    0.52}$  & $  147.14_{-    0.30-    0.59}^{+    0.30+    0.60}$  & $  147.47_{-    0.26-    0.51}^{+    0.26+    0.52}$  \\

\hline\hline    
\end{tabular}  
\caption{Observational constraints on the interacting scenario {\bf IWDM-DEq} using various datasets.}\label{tab:IWDM-DEq-1}         
\end{table*}            
\end{center}       
\endgroup 
\begin{figure*}
    \centering
\includegraphics[width=0.85\textwidth]{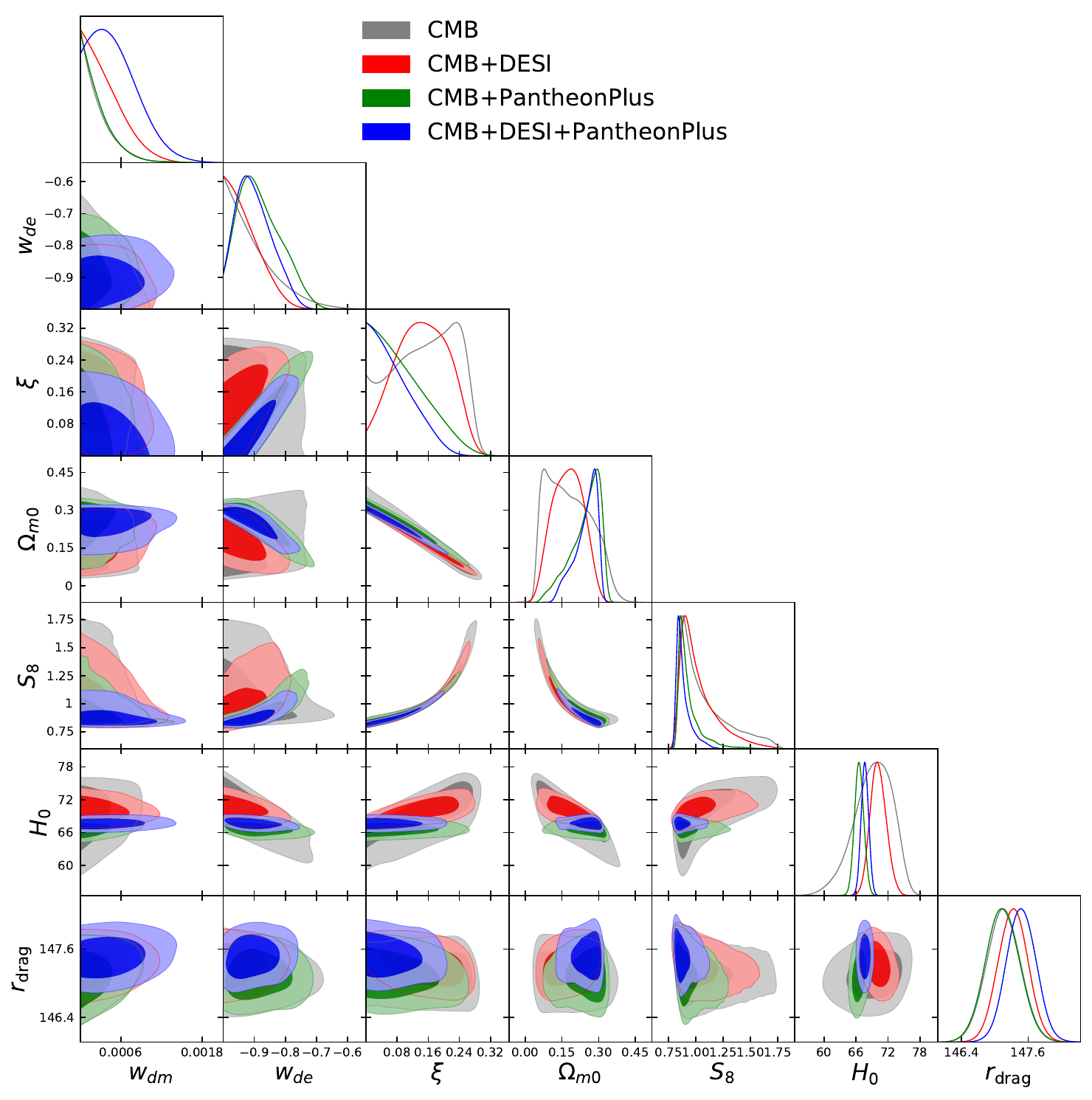}
\caption{One-dimensional marginalized posterior distributions and two-dimensional joint contours for the interacting scenario {\bf IWDM-DEq}, considering several combinations of cosmological observations.}
    \label{fig:IWDM-DEq-1}
\end{figure*}
\begingroup                                        
\squeezetable  
\begin{center}                
\begin{table*}                                            \scalebox{0.95}{   
\begin{tabular}{cccccccccccccc}                
\hline\hline      
Parameters & CMB & CMB+DESI & CMB+PantheonPlus & CMB+DESI+PantheonPlus \\ \hline

$\Omega_{\rm dm} h^2$ & $    0.136_{-    0.011-    0.014}^{+    0.007+    0.017}$ & $    0.134_{-    0.012-    0.015}^{+    0.008+    0.016}$ & $    0.1381_{-    0.0084-    0.013}^{+    0.0086+    0.013}$  & $    0.136_{-    0.006-    0.015}^{+    0.011+    0.013}$  \\

$\Omega_b h^2$ & $    0.02247_{-    0.00017-    0.00033}^{+    0.00017+    0.00034}$ & $    0.02239_{-    0.00017-    0.00033}^{+    0.00017+    0.00033}$  & $    0.02244_{-    0.00017-    0.00034}^{+    0.00017+    0.00034}$  & $    0.02236_{-    0.00017-    0.00033}^{+    0.00016+    0.00034}$  \\

$100\theta_{MC}$ & $    1.03969_{-    0.00054-    0.0011}^{+    0.00060+    0.0010}$ & $    1.03994_{-    0.00056-    0.0010}^{+    0.00057+    0.0010}$ & $    1.03957_{-    0.00053-    0.00091}^{+    0.00049+    0.00095}$  & $    1.03988_{-    0.00057-    0.00087}^{+    0.00044+    0.00099}$ \\

$\tau$ & $    0.0521_{-    0.0074-    0.015}^{+    0.0074+    0.015}$ & $    0.0554_{-    0.0081-    0.015}^{+    0.0074+    0.016}$ & $    0.0529_{-    0.0076-    0.016}^{+    0.0076+    0.016}$ & $    0.0569_{-    0.0082-    0.015}^{+    0.0075+    0.017}$ \\

$n_s$ & $    0.9680_{-    0.0051-    0.010}^{+    0.0050+    0.010}$ & $    0.9742_{-    0.0040-    0.0078}^{+    0.0040+    0.0079}$ & $    0.9674_{-    0.0049-    0.0096}^{+    0.0048+    0.0097}$ & $    0.9767_{-    0.0038-    0.0075}^{+    0.0038+    0.0076}$ \\

${\rm{ln}}(10^{10} A_s)$ & $    3.052_{-    0.015-    0.030}^{+    0.015+    0.030}$ & $    3.055_{-    0.016-    0.032}^{+    0.016+    0.033}$ & $    3.054_{-    0.016-    0.031}^{+    0.016+    0.032}$  & $    3.057_{-    0.017-    0.031}^{+    0.015+    0.034}$ \\

$w_{dm}$ & $ -0.00109_{- 0.00069- 0.0014}^{+    0.00070+    0.0013}$ & $  -0.00009_{- 0.00048-    0.00093}^{+ 0.00048+ 0.00094}$ & $ -0.00114_{-    0.00063-    0.0013}^{+    0.00064+    0.0013}$ & $    0.00035_{-    0.00044-    0.00086}^{+    0.00044+    0.00085}$  \\

$w_{de}$ & $  -1.66_{-    0.37-    0.56}^{+    0.35+    0.64}$ & $   -1.179_{-    0.074-    0.16}^{+    0.092+    0.16}$ & $   -1.087_{-    0.039-    0.086}^{+    0.060+    0.087}$ & $   -1.045_{-    0.012-    0.057}^{+    0.045+    0.045}$ \\

$\xi$ & $ > -0.049 > -0.089 $ & $ > -0.064 > -0.098 $ 
 & $ -0.053_{- 0.027}^{+    0.043}, \; > -0.102 $ & $   -0.061_{-    0.040-    0.046}^{+ 0.020+ 0.052}$ \\

$\Omega_{m0}$ & $    0.245_{-    0.088-    0.10}^{+    0.036+    0.13}$ & $    0.312_{-    0.026-    0.041}^{+    0.022+    0.045}$ & $    0.378_{-    0.023-    0.042}^{+    0.023+    0.043}$ & $    0.347_{-    0.017-    0.040}^{+    0.025+    0.035}$ \\

$\sigma_8$ & $    0.882_{-    0.096-    0.15}^{+    0.088+    0.16}$ & $    0.780_{-    0.046-    0.074}^{+    0.041+    0.075}$  & $    0.747_{-    0.037-    0.055}^{+    0.030+    0.057}$  & $    0.736_{-    0.044-    0.056}^{+    0.024+    0.068}$ \\

$H_0$ & $   82_{-   11-   18}^{+   11+   18}$ & $   71.1_{-    1.9-    3.3}^{+    1.6+    3.5}$ & $   65.32_{-    0.98-    1.9}^{+    0.99+    2.0}$ & $   67.73_{-    0.70-    1.3}^{+    0.69+    1.4}$ \\

$S_8$ & $    0.782_{-    0.043-    0.073}^{+    0.039+    0.078}$ & $    0.793_{-    0.019-    0.035}^{+    0.019+    0.036}$ 
 & $    0.838_{-    0.023-    0.045}^{+    0.023+    0.045}$ & $    0.790_{-    0.020-    0.033}^{+    0.017+    0.037}$  \\

$r_{\rm{drag}}$ & $  146.75_{-    0.37-    0.72}^{+    0.37+    0.73}$ & $  147.24_{-    0.27-    0.52}^{+    0.27+    0.54}$ & $  146.72_{-    0.34-    0.68}^{+    0.34+    0.67}$   & $  147.44_{-    0.25-    0.50-    0.65}^{+    0.25+    0.50}$ \\

\hline\hline                           
\end{tabular}  }                     
\caption{68\% and 95\% CL constraints on the free and derived parameters of the {\bf IWDM-DEp} scenario, assuming $w_{\rm dm}$ is a free parameter varying in the range $[-1, 1]$, and considering CMB from Planck 2018, CMB+DESI, CMB+PantheonPlus, and CMB+DESI+PantheonPlus.}
\label{tab:IWDM-DEp-2}                 
\end{table*}          
\end{center}              
\endgroup                                                 \begin{figure*}
    \centering
    \includegraphics[width=0.85\textwidth]{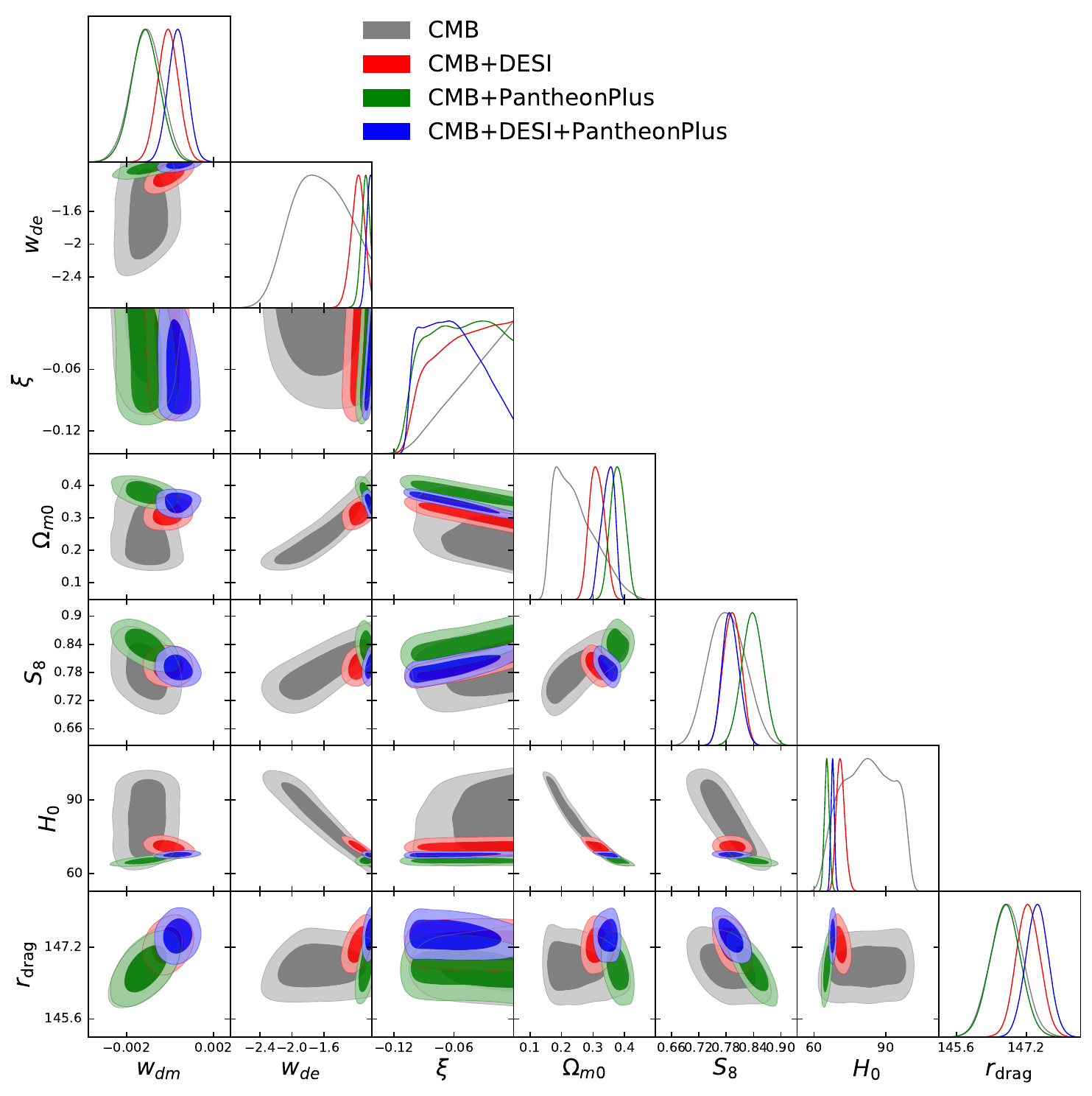}
\caption{A wider version of the {\bf IWDM-DEp} scenario, allowing $\xi$ to vary in the range $[-1, 1]$ and considering all possible combinations of the datasets.}
    \label{fig:IWDM-DEp-2}
\end{figure*}                               

\begingroup                                                                                          
\squeezetable                                                                                                                   
\begin{center}                                                                                                                  
\begin{table*}                                                                                                      \scalebox{0.95}{  
\begin{tabular}{ccccccccc}                  
\hline\hline                      
Parameters & CMB & CMB+DESI & CMB+PantheonPlus & CMB+DESI+PantheonPlus\\ \hline

$\Omega_{\rm dm} h^2$ & $  <0.082\,<0.118 $ & $    0.049_{-    0.039}^{+    0.023}\,<0.097$  & $    0.075_{-    0.018-    0.060}^{+    0.040+    0.048}$ & $    0.084_{-    0.012-    0.056}^{+    0.034+    0.039}$ \\

$\Omega_b h^2$ & $    0.02244_{-    0.00017-    0.00033}^{+    0.00017+    0.00033}$ & $    0.02238_{-    0.00017-    0.00033}^{+    0.00017+    0.00033}$ & $    0.02243_{-    0.00017-    0.00034-    0.00044}^{+    0.00017+    0.00034}$ & $    0.02235_{-    0.00017-    0.00033}^{+    0.00017+    0.00034}$ \\

$100\theta_{MC}$ & $    1.0444_{-    0.0038-    0.0043}^{+    0.0020+    0.0048}$ & $    1.0454_{-    0.0027-    0.0038}^{+    0.0022+    0.0040}$ & $    1.0434_{-    0.0027-    0.0032}^{+    0.0010+    0.0045}$ & $    1.0430_{-    0.0022-    0.0027}^{+    0.0007+    0.0040}$ \\

$\tau$ & $    0.0524_{-    0.0074-    0.015}^{+    0.0075+    0.015}$ & $    0.0560_{-    0.0083-    0.015}^{+    0.0076+    0.016}$  & $    0.0529_{-    0.0079-    0.015}^{+    0.0074+    0.016}$ & $    0.0571_{-    0.0085-    0.0158-    0.022}^{+    0.0077+    0.017}$ \\

$n_s$ & $    0.9663_{-    0.0051-    0.0099}^{+    0.0051+    0.0101}$ & $    0.9747_{-    0.0039-    0.0077}^{+    0.0040+    0.0080}$ & $    0.9668_{-    0.0049-    0.0095}^{+    0.0049+    0.0098}$ 
 & $    0.9769_{-    0.0039-    0.0076}^{+    0.0039+    0.0079}$ \\

${\rm{ln}}(10^{10} A_s)$ & $    3.054_{-    0.015-    0.031}^{+    0.015+    0.031}$ & $  3.056_{-    0.016-    0.032}^{+    0.016+    0.032}$ & $    3.054_{-    0.016-    0.031}^{+    0.016+    0.032}$  & $ 3.057_{-    0.016-    0.033}^{+    0.016+    0.035}$ \\

$w_{\rm dm}$ & $   -0.00134_{-    0.00068-    0.0014}^{+    0.00069+    0.0014}$ & $   -0.00004_{-    0.00048-    0.00092}^{+    0.00047+    0.00096}$ & $   -0.00121_{-    0.00066-    0.0013}^{+    0.00066+    0.0013}$ & $    0.00037_{-    0.00045-    0.00085}^{+    0.00044+    0.00085}$ \\

$w_{de}$ & $ < -0.864 < -0.717 $ & $ < -0.909 < -0.830 $ & $ < -0.850 < -0.754 $ & $ -0.889_{- 0.086}^{+    0.043},\;  < -0.845$  \\

$\xi$ & $ 0.19_{- 0.08-    0.19}^{+ 0.13+ 0.15}$ & $    0.180_{- 0.049- 0.13}^{+ 0.091+ 0.12}$  & $ 0.15_{-  0.11-  0.15}^{+ 0.07+ 0.15}$ & $ < 0.116 < 0.214 $ \\

$\Omega_{m0}$ & $    0.20_{-    0.15-    0.16}^{+    0.07+    0.19}$ &  $    0.147_{-    0.086-    0.11}^{+    0.045+    0.12}$ & $ 0.231_{-    0.047-    0.14}^{+    0.093+    0.11}$  & $    0.234_{-    0.029-    0.12}^{+    0.074+    0.09}$ \\

$\sigma_8$ & $    1.6_{-    1.0-    1.1}^{+    0.2+    2.1}$ &  $    1.80_{-    0.98-    1.1}^{+    0.26+    1.9}$ & $    1.28_{-    0.51-    0.6}^{+    0.06+    1.2}$ & $    1.10_{-    0.34-    0.44}^{+    0.03+    0.80}$ \\

$H_0$ & $   66.0_{-    3.3-    7.1}^{+    4.2+    6.8}$ &  $   70.5_{-    1.6-    2.9}^{+    1.5+    3.0}$ & $   65.2_{-    1.1-    1.9}^{+    1.0+    2.0}$ & $   67.64_{-    0.70-    1.4}^{+    0.71+    1.4}$ \\

$S_8$ & $    1.14_{-    0.30-    0.33}^{+    0.09+    0.55}$ &  $    1.12_{-    0.28-    0.33}^{+    0.10+    0.48}$ & $    1.03_{-    0.19-    0.23}^{+    0.04+    0.40}$ & $    0.92_{-    0.12-    0.15}^{+    0.02+    0.26}$ \\

$r_{\rm{drag}}$ & $  146.62_{-    0.37-    0.72}^{+    0.37+    0.73}$ &   $  147.28_{-    0.27-    0.52}^{+    0.27+    0.53}$ & $  146.67_{-    0.35-    0.69}^{+    0.35+    0.70}$ & $  147.46_{-    0.26-    0.51}^{+    0.26+    0.52}$  \\

\hline\hline                       
\end{tabular} } 
\caption{68\% and 95\% CL constraints on the free and derived parameters of the \textbf{IWDM-DEq} scenario assuming that $w_{\rm dm}$ is a free-to-vary parameter in $[-1, 1]$ and considering CMB from Planck 2018, CMB+DESI, CMB+PantheonPlus, and CMB+DESI+PantheonPlus.    }
\label{tab:IWDM-DEq-2}         
\end{table*}                      
\end{center}                      
\endgroup                          \begin{figure*}
    \centering
\includegraphics[width=0.85\textwidth]{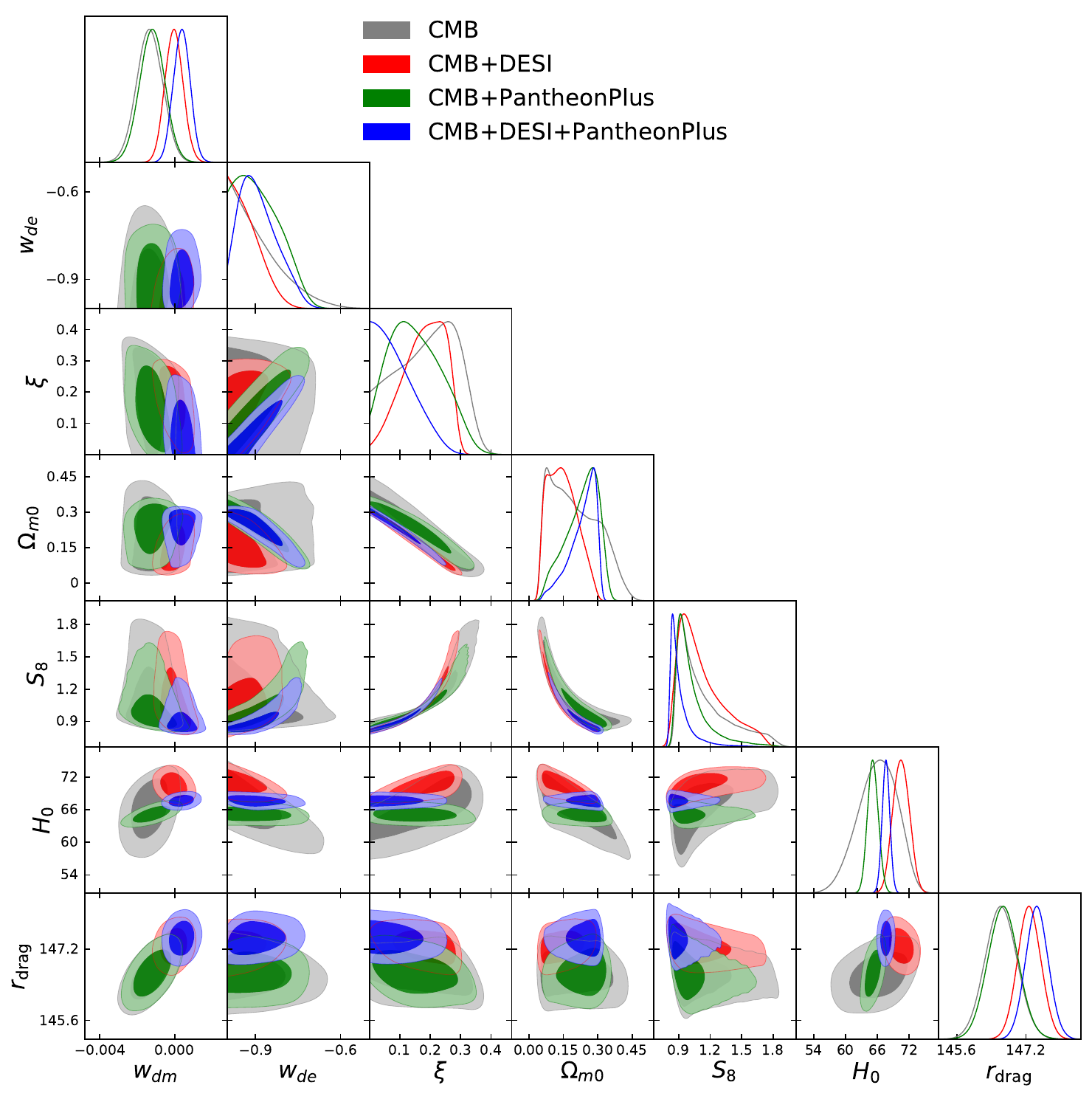}
\caption{A wider version of the {\bf IWDM-DEq} scenario, allowing $\xi$ to vary in the range $[-1, 1]$ and considering all possible combinations of the datasets.}
    \label{fig:IWDM-DEq-2}
\end{figure*}

\section{Results}
\label{sec-results}

In this section, we describe the observational constraints obtained after analyzing the interacting scenarios. In the first two subsections, we present the constraints on the interacting models under the assumption $w_{\rm dm} \in [0, 1]$, as specified in Table~\ref{tab:priors}. Subsequently, we relax this assumption by allowing $w_{\rm dm}$ to vary freely in the range $[-1, 1]$ to investigate whether observational data suggest any preference for a negative EoS for the DM component.  
Thus, the key parameters of the interacting scenarios under consideration are $w_{\rm dm}$ and $\xi$. Our primary objective is to determine whether observational evidence supports nonzero values of $w_{\rm dm}$ and $\xi$, and if so, to assess how these parameters influence the overall cosmological scenario, particularly in relation to the remaining key parameters.

\subsection{IWDM-DEp}

Table~\ref{tab:IWDM-DEp-1} and Fig.~\ref{fig:IDWM-DEp-1} summarize the observational constraints on this interacting scenario. When considering CMB data alone, no significant evidence for $w_{\rm dm} \neq 0$ or $\xi \neq 0$ is found; instead, a non-interacting cosmological scenario remains in good agreement with observations. However, an interesting result emerges: we obtain a low value for the parameter $S_8 = 0.755^{+0.034}_{-0.038}$, which significantly alleviates the $S_8$ tension between Planck~\cite{Planck:2018vyg} and cosmic shear measurements, including low-redshift lensing and galaxy cluster observations~\cite{DES:2017myr,Troster:2019ean,Heymans:2020gsg,Kobayashi:2021oud,DES:2021wwk,Miyatake:2021sdd,SimBIG:2023nol}.  
When DESI-BAO observations are incorporated alongside CMB data, the values of $w_{\rm dm}$ and $\xi$ remain similar to those in the CMB-only case (i.e., $w_{\rm dm} = 0$ and $\xi = 0$ are consistent within 68\% CL). However, unlike the CMB-only scenario, which favored a very high value of $H_0$, in this case, we obtain $H_0 = 70.6^{+1.5}_{-1.8}$ km/s/Mpc at 68\% CL along with $S_8 = 0.794^{+0.019}_{-0.020}$ at 68\% CL, continuing to provide relief for both the $H_0$ and $S_8$ tensions. 

When PantheonPlus observations are combined with CMB measurements, no evidence for $w_{\rm dm} \neq 0$ is found. However, an indication of a nonzero interaction is obtained at more than 95\% CL, with $\xi = -0.063^{+0.051}_{-0.046}$ at 95\% CL for CMB+PantheonPlus. This result comes at the expense of a higher value for the current matter density parameter, $\Omega_{m0} = 0.361^{+0.025}_{-0.019}$ at 68\% CL (CMB+PantheonPlus), due to the energy flow from DE to DM, which consequently leads to a lower value of $H_0$. Additionally, this dataset combination yields a lower value of $S_8 = 0.809^{+0.019}_{-0.021}$ at 68\% CL, compared to the Planck-based $\Lambda$CDM framework~\cite{Planck:2018vyg}, thereby mildly alleviating the $S_8$ tension.
Finally, for the combined dataset CMB+DESI+PantheonPlus, $w_{\rm dm}$ is found to be nonzero at slightly more than $1\sigma$ significance, with $w_{\rm dm} = 0.00050^{+0.00015}_{-0.00048}$ at 68\% CL (CMB+DESI+PantheonPlus), though within 95\% CL, $w_{\rm dm} = 0$ remains consistent with the data. Moreover, an interaction is detected at more than 95\% CL, with $\xi = -0.063^{+0.051}_{-0.045}$ at 95\% CL, again implying an energy flow from DE to DM. In this case, we also find a lower value of $S_8 = 0.788^{+0.016}_{-0.020}$ at 68\% CL, further alleviating the $S_8$ tension. However, the value of $H_0$ remains very similar to that of the Planck-based $\Lambda$CDM model.

\subsection{IWDM-DEq}

Table~\ref{tab:IWDM-DEq-1} and Fig.~\ref{fig:IWDM-DEq-1} summarize the constraints on this interacting scenario.  
Starting with the constraints from CMB alone, we find no evidence for $w_{\rm dm} \neq 0$. On the other hand, although $\xi$ is statistically consistent with zero, its effects on the current matter density parameter are evident, yielding a much lower value of $\Omega_{m0} = 0.180^{+0.06}_{-0.13}$ at 68\% CL compared to the standard $\Lambda$CDM framework. This result arises due to the energy transfer from the DM component to the DE component.  
As a consequence of this energy flow, a higher value of the Hubble constant is naturally obtained, with $H_0 = 69.2^{+4.3}_{-2.9}$ km/s/Mpc at 68\% CL.  
When DESI-BAO data is added to CMB, our conclusion on $w_{\rm dm}$ remains unchanged, meaning that $w_{\rm dm}$ is still consistent with zero at $1\sigma$. However, we find evidence for an interaction at more than $2\sigma$, with $\xi = 0.15 \pm 0.12$ at 95\% CL for CMB+DESI. Since $\xi > 0$, as previously stated, energy flows from DM to DE. As a result, a lower value of the matter density parameter is still observed, with $\Omega_{m0} = 0.173^{+0.066}_{-0.069}$ at 68\% CL, along with a higher value of $H_0 = 70.2^{+1.4}_{-1.6}$ km/s/Mpc at 68\% CL.  
Therefore, the combination of CMB+DESI favors an interaction in the dark sector, which helps alleviate the $H_0$ tension between Planck and SH0ES.

However, the inclusion of PantheonPlus with CMB leads to different results. In this case, no evidence for $w_{\rm dm} \neq 0$ or for an interaction (i.e., $\xi \neq 0$) is found. On the contrary, CMB+PantheonPlus reduces the mean value of $H_0$ compared to CMB alone, bringing it in line with the non-interacting $\Lambda$CDM cosmology as inferred by Planck~\cite{Planck:2018vyg}.  
The combination of CMB, DESI, and PantheonPlus provides very mild evidence for a nonzero value of $w_{\rm dm}$ ($w_{\rm dm} \neq 0$ at 68\% CL, though still consistent with zero within 95\% CL) along with evidence for an interaction at more than 95\% CL. However, in this case, the estimated value of $\Omega_{m0}$ is higher than in the CMB and CMB+DESI scenarios, leading to an $H_0$ value that is very similar to its Planck-based $\Lambda$CDM prediction~\cite{Planck:2018vyg}.

\subsection{Allowing for more freedom in the DM EoS}

In the earlier sections, we presented results under the assumption that the DM EoS remains non-negative, i.e., $w_{\rm dm} \geq 0$. While this is the most widely accepted assumption in modern cosmology, here we take a heuristic approach by allowing $w_{\rm dm}$ to vary within $[-1,1]$ to investigate whether observational data suggest deviations from the usual scenario. In other words, we permit the EoS of DM to take negative values and allow observational data to determine its preferred range. Keeping the same flat priors on all other parameters (see Table~\ref{tab:priors}), we perform the analyses within the {\bf IWDM-DEp} and {\bf IWDM-DEq} cosmologies, considering the same observational datasets.  
Tables~\ref{tab:IWDM-DEp-2} and \ref{tab:IWDM-DEq-2} present the results for these two scenarios, and Figs.~\ref{fig:IWDM-DEp-2} and \ref{fig:IWDM-DEq-2} provide the corresponding graphical representations.  

Focusing on {\bf IWDM-DEp}, we observe that for CMB alone, there is a mild preference for a negative $w_{\rm dm}$ ($w_{\rm dm} \neq 0$ at slightly more than 68\% CL, though still consistent with zero within 95\% CL). This trend is also supported by CMB+PantheonPlus, while for CMB+DESI, $w_{\rm dm}$ remains consistent with zero at 68\% CL. However, for the combined dataset CMB+DESI+PantheonPlus, the preferred values of $w_{\rm dm}$ shift to the positive range (though still consistent with zero at 68\% CL), along with a preference for an interaction at more than 95\% CL. Thus, a mild deviation from $w_{\rm dm} = 0$ is suggested across all data combinations. Moreover, except for the CMB+PantheonPlus combination, the other three cases yield a lower value of $S_8$, alleviating the $S_8$ tension between Planck and cosmic shear measurements.  

For the other scenario, {\bf IWDM-DEq}, we obtain similar results as in {\bf IWDM-DEp}. Specifically, for CMB and CMB+PantheonPlus, $w_{\rm dm}$ remains in the negative range at slightly more than 68\% CL but is still consistent with zero within 95\% CL (or within 68\% CL for CMB+DESI). For the final case, CMB+DESI+PantheonPlus, the mean value of $w_{\rm dm}$ is positive, though still consistent with zero at 68\% CL. Additionally, CMB+DESI+PantheonPlus exhibits a preference for an interaction at more than 95\% CL.

\section{Summary and Conclusions}
\label{sec-summary}

An interacting DM$-$DE scenario, where energy exchange between these two dark components is allowed, is the central theme of this work. Models in this category have received significant attention in the cosmology community due to their potential to address key cosmic puzzles. From the cosmic coincidence problem (the {\it ``why now''} problem in cosmology) to current cosmological tensions indicated by various astronomical observations, interacting cosmological scenarios have demonstrated their ability to alleviate such issues.  
The interaction function, $Q$, which appears in the conservation equations of DM and DE (see Eqs.~(\ref{cons-de}) and (\ref{cons-wdm})), plays a crucial role in the dynamics of the universe, as it modifies the expansion history at both background and perturbation levels. However, beyond the interaction function itself, the nature of the dark components is also a key ingredient in the interacting dynamics. For instance, the DE involved in the interaction process could be vacuum energy decaying into DM, or it could represent a DE fluid with a constant or dynamical equation of state. Similarly, the DM component could be either cold or non-cold.  
Nevertheless, in almost all interacting DM-DE models, the pressure of the DM component is assumed to be zero. This assumption of cold DM in the context of DM-DE interactions is primarily motivated by the present-day abundance of CDM. However, since the fundamental nature of DM and DE remains unknown, there is no a priori reason to impose additional constraints on any cosmic component. In fact, the current abundance of CDM in the dark sector might be the result of decaying dark matter that was non-cold at earlier times.

In the present work, we have raised the above question and investigated an interacting DM-DE scenario in which the EoS parameters of DM and DE are constant. Considering the large-scale stability of this interacting scenario, which depends on both $w_{\rm de}$ and the coupling parameter $\xi$ of the interaction function, we have divided the parameter space into two regions:  

\begin{itemize}
    \item[(i)] $w_{\rm de} < -1$, $\xi \leq 0$ \quad\quad   (labeled as {\bf IWDM-DEp})  
    \item[(ii)] $w_{\rm de} > -1$, $\xi \geq 0$ \quad\quad   (labeled as {\bf IWDM-DEq})  
\end{itemize}  

We constrained these two scenarios using CMB observations from Planck 2018, BAO from DESI measurements, and the PantheonPlus sample of SNe Ia. We explored two different cases: one in which $w_{\rm dm}$ is restricted to non-negative values (Tables~\ref{tab:IWDM-DEp-1} and \ref{tab:IWDM-DEq-1}), and another in which $w_{\rm dm}$ is allowed to take both negative and positive values (Tables~\ref{tab:IWDM-DEp-2} and \ref{tab:IWDM-DEq-2}). Our findings reveal some interesting results:

\begin{itemize}

\item When $w_{\rm dm}$ is restricted to positive values, we find that for both {\bf IWDM-DEp} and {\bf IWDM-DEq}, an indication of $w_{\rm dm} \neq 0$ at slightly more than 68\% CL is obtained only for CMB+DESI+PantheonPlus. Additionally, in the {\bf IWDM-DEp} scenario, evidence for an interaction at more than 95\% CL is found for CMB+PantheonPlus and CMB+DESI+PantheonPlus, while for {\bf IWDM-DEq}, interaction is detected at more than 95\% CL only for CMB+DESI. Furthermore, the {\bf IWDM-DEp} cosmology alleviates the $S_8$ tension in all cases explored in this work, whereas the {\bf IWDM-DEq} scenario does not. Notably, for the {\bf IWDM-DEp} case, the CMB+DESI combination is particularly interesting as it also alleviates the $H_0$ tension.  

\item When $w_{\rm dm}$ is allowed to vary freely in the range $[-1, 1]$, we find that for both CMB and CMB+PantheonPlus, the mean value of $w_{\rm dm}$ is negative regardless of the cosmological scenario ({\bf IWDM-DEp} or {\bf IWDM-DEq}), with $w_{\rm dm} \neq 0$ at slightly more than 68\% CL (though consistent with zero at 68\% CL for CMB+DESI). In the combined analysis CMB+DESI+PantheonPlus, the mean value of $w_{\rm dm}$ is positive but remains consistent with zero within 68\% CL. Moreover, in the \textbf{IWDM-DEp} scenario, evidence for an interaction is found for CMB+PantheonPlus (at more than 68\% CL), and CMB+DESI+PantheonPlus (at more than 95\% CL), while for \textbf{IWDM-DEq}, interaction is detected at more than 68\% CL for CMB and CMB+PantheonPlus, but for CMB+DESI, this evidence is found at more than 95\% CL. 
Additionally, we find that, except for CMB+PantheonPlus, the {\bf IWDM-DEp} scenario continues to weaken the $S_8$ tension between the CMB-based $\Lambda$CDM prediction and cosmic shear measurements. 

\end{itemize}

On the basis of the outcomes, there is no doubt that the interacting scenario \textbf{IWDM-DEp} (energy transfer occurs from DE to DM), is appealing specifically because of its ability to alleviate the the $S_8$ tension, however, considering the evidence of interaction in \textbf{IWDM-DEq} (energy flow occurs from DM to DE) which is pronounced when $w_{\rm dm}$ is allowed to vary in a wide region,   one may argue that the present-day abundance of CDM might be the result of a decaying non-cold DM component transitioning into DE due to an interaction between the two. Although such scenario will be physically realistic if a dynamical $w_{\rm dm}$ is considered instead of its constant nature, because, with such  dynamical EoS of DM, it might be possible to realize the transition of $w_{\rm dm}$ from its past non-cold nature to present cold nature, and this needs further investigations as a future work.  Our analysis of the latest cosmological data suggests that a small fraction of non-cold DM may still exist in the present universe. This is an intriguing result, as it could provide insights into small-scale structures and galaxy properties that are in tension with a purely CDM-dominated scenario.

\section{Acknowledgments}
We thank the referee very much for his/her important comments that improved the quality of the manuscript. 
WY has been is supported by the National Natural Science Foundation of China under Grants No. 12175096, and Liaoning Revitalization Talents Program under Grant no. XLYC1907098. SP acknowledges the financial support from  the Department of Science and Technology (DST), Govt. of India under the Scheme  ``Fund for Improvement of S\&T Infrastructure (FIST)'' [File No. SR/FST/MS-I/2019/41].   EDV is supported by a Royal Society Dorothy Hodgkin Research Fellowship. OM acknowledges the financial support from the MCIU with funding from the European Union NextGenerationEU (PRTR-C17.I01) and Generalitat Valenciana (ASFAE/2022/020). This work has been supported by the Spanish grants  PID2023-148162NB-C22 and PID2020-113644GB-I00 and by the European ITN project HIDDeN (H2020-MSCA-ITN-2019/860881-HIDDeN) and SE project ASYMMETRY (HORIZON-MSCA-2021-SE-01/101086085-ASYMMETRY) and well as by the Generalitat Valenciana grant CIPROM/2022/69. D.F.M. acknowledges support from the Research Council of Norway and UNINETT Sigma2 -- the National Infrastructure for High Performance Computing and Data Storage in Norway. This article is based upon work from COST Action CA21136 Addressing observational tensions in cosmology with systematics and fundamental physics (CosmoVerse) supported by COST (European Cooperation in Science and Technology).



\bibliography{biblio}
\end{document}